%% file: main.tex
\documentclass[conference,compsoc]{IEEEtran}

\usepackage{boldline}
\usepackage{booktabs}
\usepackage{graphicx}
\usepackage{listings}
\usepackage{makecell}
\usepackage{mathtools}
\usepackage{multirow}
\usepackage{pifont}
\usepackage{subfig}
\usepackage{todonotes}
\usepackage{xspace}
\usepackage{hyperref}

\newcommand{\refappendix}[1]{\hyperref[#1]{Appendix~\ref*{#1}}}

\usepackage{xurl}

\usepackage{breakurl}
\usepackage{microtype}
\usepackage{extarrows}
\usepackage{textcomp}
\usepackage{xparse}   
\usepackage{amsmath}
\usepackage{bigdelim}
\usepackage[linesnumbered,ruled,vlined]{algorithm2e}
\usepackage{color}
\usepackage{booktabs}
\usepackage{tabularx}
\usepackage[T1]{fontenc}
\usepackage{tikz}

\ifCLASSOPTIONcompsoc
  \usepackage[nocompress]{cite}
\else
  \usepackage{cite}
\fi
\def\ptitle{\sysname: Precise Call Graph Construction for Split-Phase
Applications using Dynamic Seeding}
\def\pkeywords{}

\newcommand{\cmark}{\ding{51}}%
\newcommand{\xmark}{\ding{55}}%

\newcommand{\sysname}{PhaseSeed\xspace}

\newcommand*\colourcheck[1]{%
  \expandafter\newcommand\csname #1check\endcsname{\textcolor{#1}{\ding{52}}}%
  }
\colourcheck{black}
\colourcheck{green}
\colourcheck{red}

\newcolumntype{P}[1]{>{\raggedright\arraybackslash}p{#1}}
\newcolumntype{E}[1]{>{\raggedleft\arraybackslash}p{#1}}
\newcolumntype{C}[1]{>{\centering\arraybackslash}p{#1}}

\hypersetup{
	colorlinks	= true,
	citecolor	= black,
	filecolor	= black,
	linkcolor	= black,
	urlcolor	= black,
	pagebackref	= false,
	pdftitle	= {\ptitle},
	pdfkeywords	= {\pkeywords}
}

\definecolor{mGreen}{rgb}{0,0.6,0}
\definecolor{mGray}{rgb}{0.5,0.5,0.5}
\definecolor{mPurple}{rgb}{0.58,0,0.82}
\definecolor{backgroundColour}{rgb}{0.95,0.95,0.92}

\lstdefinestyle{CStyle}{
    backgroundcolor=\color{backgroundColour},   
    commentstyle=\color{mGreen},
    keywordstyle=\color{magenta},
    numberstyle=\tiny\color{mGray},
    stringstyle=\color{mPurple},
    basicstyle=\footnotesize\ttfamily,
    breakatwhitespace=false,         
    breaklines=true,                 
    captionpos=b,                    
    keepspaces=true,                 
    numbers=left,                    
    numbersep=5pt,                  
    showspaces=false,                
    showstringspaces=false,
    showtabs=false,                  
    tabsize=2,
    language=C
}
\newcommand{\hamed}[1]{\ding{110}\ding{43}\textcolor{brown}{HG: #1}}
\newcommand{\mikepo}[1]{\ding{110}\ding{43}\textcolor{cyan}{MP: #1}}
\newcommand{\tpalit}[1]{\ding{110}\ding{43}\textcolor{blue}{TP: #1}}
\newcommand{\sm}[1]{\ding{110}\ding{43}\textcolor{purple}{SM: #1}}

\newcommand{\squishenumerate}{
   \begin{enumerate}
       { \setlength{\itemsep}{0pt}  \setlength{\parsep}{3pt}
      \setlength{\topsep}{3pt}       \setlength{\partopsep}{0pt}
      \setlength{\leftmargin}{2em} \setlength{\labelwidth}{1em}
      \setlength{\labelsep}{0.5em} } }

\newcommand{\squishenumerateend}{
    \end{enumerate}  }

\newcommand{\squishlist}{
   \begin{list}{$\bullet$}
    { \setlength{\itemsep}{0pt}      \setlength{\parsep}{2pt}
      \setlength{\topsep}{3pt}       \setlength{\partopsep}{0pt}
      \setlength{\leftmargin}{1em} \setlength{\labelwidth}{0.6em}
      \setlength{\labelsep}{0.5em} } }

\newcommand{\squishlisttwo}{
   \begin{list}{$\bullet$}
    { \setlength{\itemsep}{0pt}    \setlength{\parsep}{0pt}
      \setlength{\topsep}{0pt}     \setlength{\partopsep}{0pt}
      \setlength{\leftmargin}{2em} \setlength{\labelwidth}{1.5em}
      \setlength{\labelsep}{0.5em} } }

\newcommand{\squishend}{
    \end{list}  }

\newcommand\para[1]{\vspace{0.5em} \noindent \textbf{#1}}
\newcommand\parasmall[1]{\vspace{0.2em} \noindent \textbf{#1}}

\makeatletter
\begin{document}

\title{\ptitle}

\author{Anonymous Submission}


\author{
{\rm Tapti Palit}\\
UC Davis\\
tpalit@ucdavis.edu
\and
{\rm Seyedhamed Ghavamnia}\\
Bloomberg\\
sghavamnia@bloomberg.net
\and
\and
{\rm Michalis Polychronakis}\\
Stony Brook University\\
mikepo@cs.stonybrook.edu
}

\definecolor{mygray}{rgb}{0.6,0.6,0.6}
\definecolor{commentcolor}{RGB}{130,100,0}
\definecolor{annotationcolor}{HTML}{cc3300}

\lstdefinestyle{myC}{
    language    = C,
    basicstyle =\linespread{0.8}\ttfamily\small,
    numbers = left,
    numbersep = 5pt,
  xleftmargin = 12pt,
    numberstyle = \scriptsize\color{mygray},
    breaklines = true,
    captionpos = b,
    keywordstyle = \color{blue},
    stringstyle = \color{red},
    commentstyle = \color{magenta},
    showstringspaces=false,
    keywordstyle=[2]\color{annotationcolor},
    keywords=[2]{mark_sensitive},
}

\maketitle
\input{abstract}
\pagestyle{plain}

\input{new_intro}
\input{bgmotivation}
\input{approach}

\input{design}
\input{impl}
\input{usecases}
\input{eval}

\input{discussion}

\input{related}
\input{conclusion}

\bibliographystyle{plain}
\bibliography{main}

\input{appendix}

\end{document}

%% file: abstract.tex
\begin{abstract}
Precise and sound
call graph construction is crucial for many software security
mechanisms. 
Unfortunately, traditional static pointer analysis techniques
used to generate application call graphs suffer from imprecision.
These techniques are
agnostic to the application's architecture and are
designed for broad applicability.
To mitigate this precision problem, we propose
\sysname, a novel technique that improves the accuracy of pointer analysis
for \emph{split-phase} applications, which
have distinct initialization and processing phases.
\sysname analyzes the initialization
phase dynamically, collecting the points-to 
relationships established at runtime. At the end
of the initialization phase, it then \emph{seeds} this information
to a static analysis stage that performs pointer analysis
for all code that stays in scope during the processing phase, improving precision.
Our observations show that, given the same
runtime configuration options, the points-to relationships established
during the initialization phase remain constant across multiple
runs. Therefore, \sysname is sound with respect to 
a given initial configuration.
%
%
We apply \sysname to three security mechanisms: control flow 
integrity (CFI), software debloating, and system call filtering.
\sysname provides up to 92.6\% precision improvement for
CFI compared to 
static
call graph construction techniques, and 
filters nine additional security-critical system calls
when used to generate Seccomp profiles.
\end{abstract}

%% file: new_intro.tex
\section{Introduction}
\label{sec:intro}

Application call graph construction is essential for multiple 
security
defense mechanisms, including control flow integrity (CFI), software
debloating, and system call filtering. 
Indirect control flow transfers
require the use of static pointer analysis techniques to 
resolve their targets. Therefore, the 
security guarantees of these defense mechanisms rely
on the accuracy of the underlying
static pointer analysis technique.
For example, CFI
restricts
the targets of indirect function calls to precomputed functions.
The effectiveness of a CFI 
mechanism depends on the 
\emph{precision} of this restriction---the \emph{fewer}
targets permitted at an indirect call site, the higher the security guarantees.
On the other hand, to ensure correct application execution, 
CFI must be \emph{sound} and 
never restrict 
valid indirect function calls. Precise
and sound 
pointer analysis techniques
are thus crucial for the effectiveness 
and correctness of security defense mechanisms.

Despite decades of research, 
achieving high precision while maintaining soundness
remains challenging for static 
pointer analysis techniques~\cite{hind,pasurvey}.
The traditional approach for improving the precision of pointer analysis is by  
increasing its \emph{sensitivity} 
and incorporating additional program information
For instance, context-sensitive pointer
analysis~\cite{efficientcontextsen,sridharanrefinement,
levelbylevel} treats every
invocation of a function as distinct, while  path-sensitive 
analysis~\cite{livshits2003tracking, sui2011spas} analyzes every 
\emph{branch} of an \texttt{if} statement independently of the other. 
Increasing
the sensitivity of pointer analysis, however,
also increases the complexity of the analysis,
limiting the scalability of these techniques to only small applications.
Consequently, many software defenses~\cite{temporal, shard, typro, 
wheredoesitgo, li2023hybrid,lu2023practical,Tongpex2019} resort to using the type information 
of function call arguments
to improve call graph precision by filtering
potential targets of indirect call sites.
However,
as we demonstrate in \autoref{bg:typepunning},
\emph{statically} deriving 
accurate type information
for all objects in C/C++ codebases is error-prone and results
in unsound call graphs for type-based CFI approaches,
such as Clang-CFI~\cite{li2020finding, clang-cfi}. 

Another promising direction is to 
augment static analysis with 
dynamic analysis~\cite{shootingheap,heapsdont,
hybrid}
by running the application with certain inputs and reusing
the dynamically observed points-to relationships to improve precision.
However, naively combining static and
dynamic analysis results and excluding all dynamically
executed code from static analysis can lead to unsound results. 
The dynamically analyzed functions can be
re-invoked with different arguments from calling contexts \emph{unseen} 
during the training phase, resulting in new points-to relationships that
are not captured. 
While such techniques are suitable for
\emph{best effort} approaches such as software testing,
which can tolerate
unsoundness,
they cannot be applied to 
software defense mechanisms where soundness is critical.

To mitigate these challenges,
we propose \sysname, a novel technique for
improving the precision of call graph extraction for
\emph{split-phase} applications. 
Split-phase
applications~\cite{temporal, syspart} consist of a 
distinct \emph{initialization} stage followed by a \emph{processing}
stage. 
During initialization,  the application
instantiates various objects and function pointers that persist
through its lifetime, whereas during the 
processing phase, the application serves various types of
requests.
Unlike traditional
call graph construction
approaches that are agnostic to the application's architecture,
\sysname leverages the split-phase nature of certain applications to 
improve precision and ensure soundness. 


Instead of naively
combining static and dynamic analysis results, 
\sysname dynamically executes only the initialization phase,
while statically analyzing the processing phase,
\emph{seeding} the static
analysis phase with the dynamically resolved pointer targets of the
executed paths. Once the dynamic results are seeded, the rest of the static
pointer analysis proceeds unmodified. Unlike previous techniques
that unconditionally exclude the dynamically executed code from the 
static analysis~\cite{shootingheap, heapsdont, hybrid},
\sysname ensures that dynamically executed
code is excluded from the static analysis \emph{only} if it is 
\emph{guaranteed} to be inaccessible
from any remaining parts of the code that were not executed
during the dynamic analysis.
Consequently,
the resulting analysis is always sound with respect
to the provided runtime configuration, and
retains the precision improvements provided by
dynamic analysis, whenever possible.

During the initialization phase, 
the application 
instantiates various objects and
function pointers that persist through the application's lifetime. 
Unlike the processing phase, which shows significant 
variation depending on the request types and the application state, 
as we discuss in \autoref{sec:eval}, 
the points-to relationships established
during the initialization phase are fully deterministic 
given
an initial runtime configuration and do not
depend on user inputs. Moreover, this
initial runtime configuration can also potentially \emph{disable}
certain options, which allows the analysis to completely
\emph{remove} the corresponding code from the analysis scope.

Dynamically analyzing
an application's initialization also provides
complete type information for the heap objects allocated
during this phase. 
Modern applications
rely on complex memory allocation paradigms (e.g., memory pools, heap allocation
wrappers) that obfuscate the flow of \emph{type information} 
to heap allocation sites,
making the static derivation of type information for these objects challenging.
During dynamic execution, the obfuscated data flows corresponding
to these complex memory allocation paradigms are easily
resolved, and the type information is easily determined. 
Furthermore, \sysname uses this type information in
a \emph{strictly} sound manner, instead of using it to filter function targets
at indirect call sites. 

\sysname requires the programmer 
to just annotate the function that denotes the start of the 
processing phase and to provide the initial runtime configuration to launch the application.
Given this information, 
\sysname 
dynamically analyzes the application's initialization
phase till the transition point is reached. Then, it 
inspects
the application's memory state to collect the points-to relationships
and type information for all in-memory objects. By performing an iterative
analysis of the functions reachable from the transition point, \sysname
automatically and soundly 
derives the functions that may be accessed by the
processing phase. Finally, \sysname 
performs static analysis on these
processing phase functions after \emph{seeding} the analysis
with the points-to relationships captured during dynamic analysis.


To demonstrate the benefits of \sysname's improved precision
we apply it to three security mechanisms:
control flow integrity, software debloating, and system call filtering.
Our results show that \sysname 
improves the precision of CFI compared to standard static
pointer analysis techniques by up to 92.6\%, facilitates the removal of more code for software
debloating, and can filter nine additional system calls 
for the applications
in our evaluation set.


In summary, our main contributions include:
\squishlist
    \item We present \sysname, a sound call graph extraction approach that improves the precision of pointer
        analysis by \emph{seeding} pointer relationships obtained dynamically.
    \item We implemented a method for dynamically deriving heap object types, 
        which is sound even in
        the presence of complex programming patterns that
        obfuscate type information.
    \item We apply \sysname to improve the precision of
        CFI, software debloating, and system call filtering, demonstrating up to 92.6\% precision
        improvement compared to state-of-the-art static call graph 
        construction techniques.
\squishend

%% file: bgmotivation.tex
\section{Background}
\label{sec:background}

\subsection{Static Pointer Analysis}
\label{bg:pta}

Static pointer analysis is a technique for resolving the points-to
targets of all pointers in an application at compilation time. 
Static pointer analysis consists of two stages. 
The first stage
collects all pointer operations in the application, along with
their operands, and stores them in a \emph{constraint
graph}. The second stage solves these relationships according to 
certain \emph{constraint resolution rules}. The constraint
resolution rules for the commonly used Andersen's
pointer analysis~\cite{andersen} are specified in \refappendix{app:cons}. 

Various works~\cite{steens, andersenwave, cycleopt, 
sui2011spas, lei2019fast, wilson1995efficient} have
attempted to improve both the precision and
scalability of pointer analysis algorithms.
However, achieving sound, high-precision results within reasonable
analysis time remains a challenge.
The fundamental reason for this imprecision is that due to the
dependencies on runtime input, it is impossible to 
reason about pointer relationships with full precision statically~\cite{hind}. 
Moreover, decisions such as whether to
analyze each invocation of a function distinctly (context-sensitivity), and
whether to
consider the order of the program statements (flow-sensitivity), also impact
the precision of the points-to results. 

Indeed,
context and flow sensitivity improves the precision
of the pointer analysis. However, maintaining 
the calling context and flow
information for complex code involving arbitrarily deep
function calls and deeply nested loops and branches
significantly increases the 
number of constraints in the constraint graph, and thus reduces 
the scalability of the analysis.
Therefore, many pointer analysis clients~\cite{temporal, c2c, muzz}
opt to use a context- and flow-insensitive pointer analysis instead, thus
sacrificing precision.

\subsection{Type Ambiguity in C/C++}
\label{bg:typepunning}
Accurate type information
can assist in pointer analysis. The type information
is trivial to derive for stack and global variables
because the type information 
is embedded in the variable definition, in the
case of these variables.
However, statically deriving the accurate type
information for heap objects 
is challenging. 

\para{Type Punning}
C/C++ does not enforce type safety. Type-casting pointers of various
types to and from \texttt{void*} is commonly used in most C/C++ codebases.
Moreover, some C applications emulate object-oriented features such as
type inheritance and polymorphism
using a technique called \emph{type punning}. 
Using this technique, the programmer can create a parent-child relationship
between two \texttt{struct} types
by embedding the \emph{parent} struct
type as the first field of the \emph{child} struct type. Alternately,
the child type could simply be defined 
to contain the same fields
as its parent type as its initial subset of fields. 
This allows
a pointer to the parent struct type to safely access 
an object of the child type as long as it only
operates on the embedded parent type fields, facilitating
the implementation of polymorphic functions in C.

\begin{figure}[t]
    \centering
    \includegraphics[trim={0cm 3cm 0cm 0cm},scale=0.33]{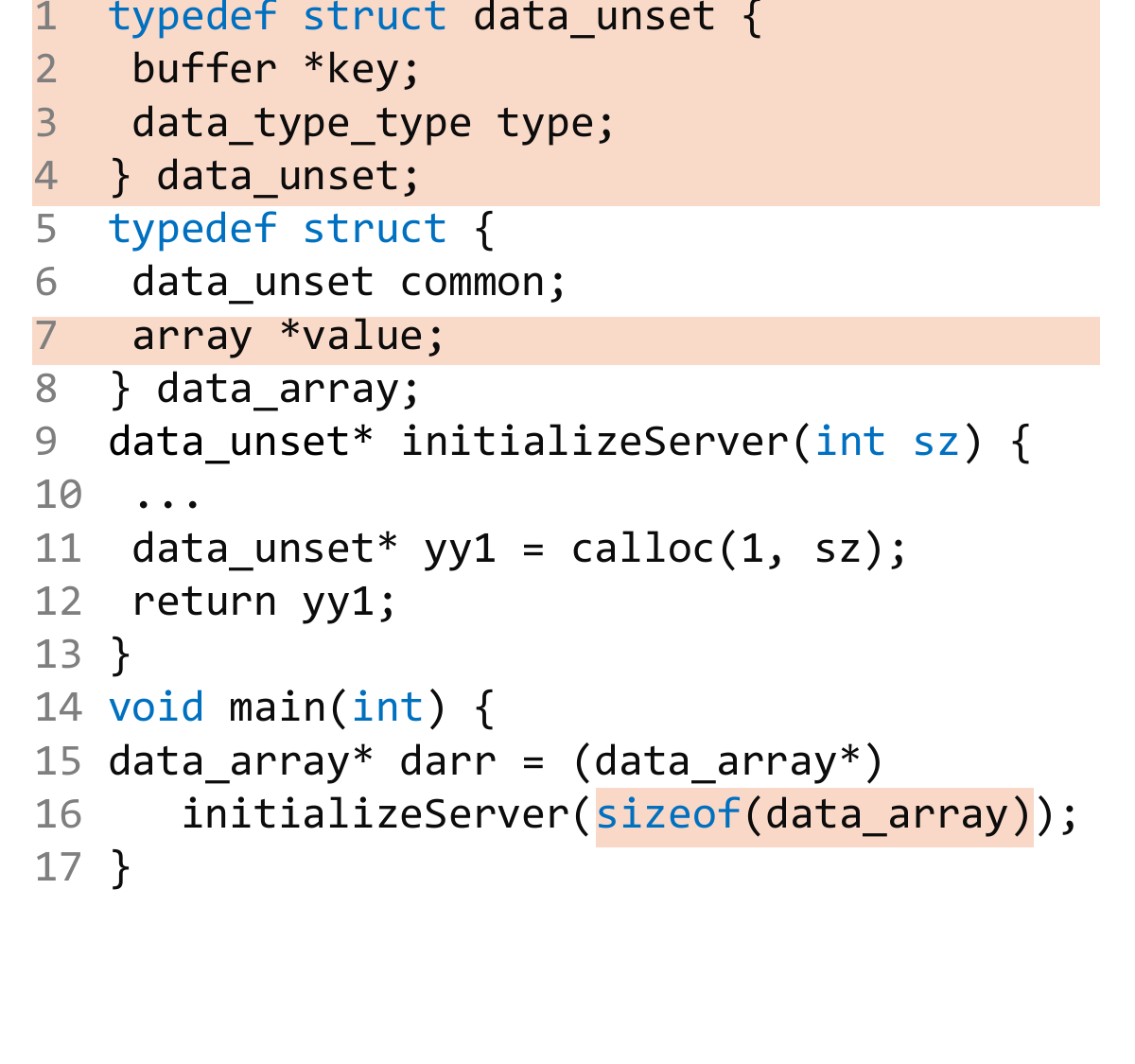}
    \caption{Type-punning in the code of Lighttpd to implement inheritance and polymorphism. 
    }
    \label{fig:typepunning}
\end{figure}

For example,
consider the simplified code snippet from the codebase 
of the popular Lighttpd web server
shown in Figure~\ref{fig:typepunning}. The type \texttt{data\_array} 
\emph{inherits} \texttt{data\_unset} by embedding a field of type
\texttt{data\_unset}.
In the function \texttt{initializeServer}, the pointer \texttt{yy1} (line 11)
of type \texttt{data\_unset} is used to initialize a heap object of type
\texttt{data\_array}. 
The object is cast
to the \emph{actual} 
\texttt{data\_array} type 
on Line 16, when the function returns.

Heap object type information is often derived by 
performing a data flow analysis from the heap 
allocation call site (e.g., \texttt{malloc}) to 
the pointer where it is stored (e.g., \texttt{data\_unset* yy1})) and tracking
every type-cast operation along the path. In the case of this example,  
however, such an analysis will only report that the \emph{partial} 
object type,
\texttt{data\_unset}, instead of the \emph{full} type, \texttt{data\_array}.
Note that while in this sample code snippet, it is feasible to 
perform an interprocedural data flow analysis to identify 
such type-casting and derive
the true types of heap objects, in real, complex codebases
the heap allocation site 
and the type-casting statement
are often separated by thousands of lines of code
across multiple functions, thus making a simple static data flow 
analysis ineffective.

\para{Heap Allocation Wrappers}
Many modern server applications use heap allocation
wrappers to perform custom checks and record-keeping 
of each heap allocation. Because multiple heap allocations
are often performed by the same heap allocation wrapper, it is no longer
possible to \emph{statically} assign a single type to the heap object
created by this pattern without employing expensive context-sensitive
analysis. Moreover, in many cases, such heap allocation wrappers 
are invoked via function pointers. To derive the type information
of heap objects allocated via indirect function calls to such heap
allocation wrappers, the analysis must first 
resolve these function pointers. These challenges
further
complicate the accurate recovery of type information for heap
objects.


\para{Impact on Precision Improvements}
Recent works~\cite{shard, temporal, wheredoesitgo, typro} have
proposed using the argument types at an indirect call site to
\emph{filter} potential function targets reported by the 
static pointer analysis.
Type ambiguity arising from type punning and heap allocation wrappers 
complicate the use of argument types to improve the
precision of the call graph construction.
With type punning, the same object can be treated as multiple
different types, depending on the emulated inheritance pattern.
Therefore, a naive precision improvement technique that does not
accurately derive these emulated type hierarchies and
simply removes functions with mismatched argument types would
therefore result 
in unsoundness. Similarly, when heap allocation wrappers
are used, the same \texttt{malloc} program statement can allocate
objects of different types and the argument type matching mechanism
must be aware of these different types allocated by
the same program statement.

To avoid these pitfalls of using statically derived type information, 
\sysname restricts the use of type information \emph{only}
to objects for which it has derived complete type information 
through dynamic execution. 

%% file: approach.tex
\section{Motivating Example}
\label{sec:motivation}

\begin{figure*}[t]
    \centering
    \includegraphics[width=\textwidth]{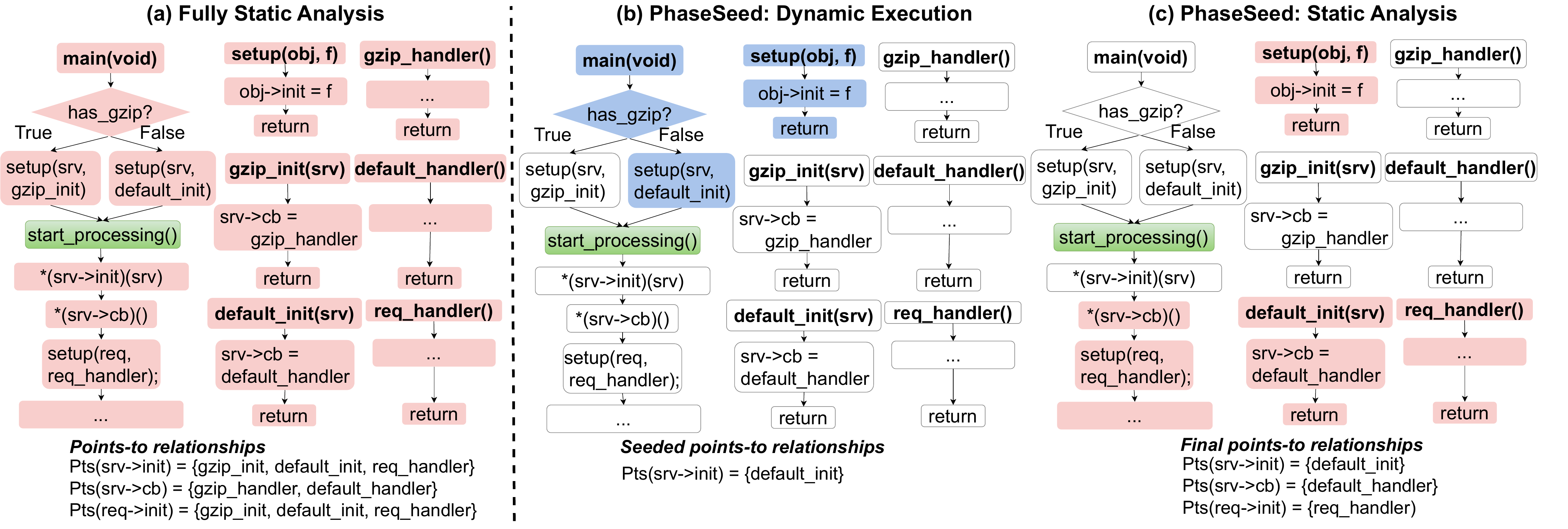}
    \caption{A fully static analysis approach compared to \sysname. 
    The \texttt{start\_processing} annotation indicates the
    transition between the initialization and processing phase 
    for \sysname. Colored boxes show the statements operated on by any given stage.
    The fully static approach derives multiple spurious points-to
    relationships due to imprecision, while \sysname's dynamic seeded approach
    can precisely derive the points-to relationships for all pointers in the
    sample code. 
   }
    \label{fig:approach}
\end{figure*}

\autoref{fig:approach} shows the control
flow graph of a simple server with
a configuration that enables \texttt{gzip} compression.
At startup, the server first checks if the 
\texttt{gzip} option is enabled or not, and 
initializes \texttt{srv->init} to 
\texttt{gzip\_init} or \texttt{default\_init}
using the \texttt{setup} function. The subsequent
\texttt{start\_processing} annotation indicates
the beginning of the processing phase. In the
processing phase, the server makes two indirect
calls through
\texttt{srv->init} and
\texttt{srv->cb}, before invoking the \texttt{setup}
function again to initialize the \texttt{req->init}
function pointer to \texttt{req\_handler}.



A fully static context-insensitive pointer analysis
technique would result in the addition of all three
callback
functions \texttt{request\_handler}, \texttt{gzip\_init}, and 
\texttt{default\_init} to the points-to sets for 
both \texttt{srv->init} 
and \texttt{request->init},
because the same function \texttt{setup} is used
to initialize all three callbacks. 
Moreover, even if the initial configuration
disables the \texttt{gzip} option, static analysis would 
still include the \texttt{gzip\_handler}
function to the points-to sets of pointers \texttt{srv->init}
and \texttt{req->init} because it does not have access to 
the runtime information necessary 
to determine that the \texttt{has\_gzip} check failed
and that only the \emph{false} branch
of the \texttt{if} statement is executed.

Dynamically executing the initialization phase
and seeding the resulting points-to relationships
into the static analysis phase mitigates the imprecision caused 
due to multiple
sources. These sources fall under the three major categories 
which we discuss below.

\subsection{Unreachable Initialization Code}
\label{sec:unreachable}
Server applications typically contain multiple
configuration options that can enable or disable
various features~\cite{c2c}. Dynamically
executing the initialization phase allows \sysname
to exclude any initialization phase code that is unreachable
under the given runtime configuration. If this
initialization code sets up callbacks via function
pointers, then excluding this code allows us 
to exclude the function targets 
that are unreachable under the given runtime configuration,
thereby improving the precision of the call graph.

For example, in the case of Figure~\ref{fig:approach},
if the initial configuration disables \emph{gzip}, then the \texttt{true}
branch of the \texttt{has\_gzip} check will not be executed, and
the dynamic execution will resolve
\texttt{srv->init} to point only to \texttt{default\_init}.
Since \texttt{gzip\_init} is not 
accessed from the processing phase \emph{at all}, 
\sysname's static analysis
will also skip analyzing it. 
Note that we exclude a function from the static analysis only if
we can guarantee that it will not be accessed from 
the processing phase. To achieve this, we use 
a conservative and sound analysis, discussed in detail in 
Section~\ref{design:partition}.

On the other hand, if a function is accessed
from the initialization phase and is also deemed accessible from the
processing phase, then we reanalyze it during static analysis. For example,
in Figure~\ref{fig:approach},
the function \texttt{setup} is invoked from both the initialization
and the processing phase. Therefore, \sysname's 
static analysis phase will reanalyze it to ensure soundness.

\subsection{Fully Sensitive Dynamic Execution}
\label{sec:fullysensitive}
As discussed in Section~\ref{bg:pta},
performing full context, flow, and
path-sensitive analysis statically 
poses a huge analysis cost. Dynamically executing
the initialization phase allows \sysname
to perform a full context, flow, and path-sensitive 
analysis for the initialization phase
code, even if the subsequent static analysis 
is context, flow, and path insensitive. Because
typical
servers are initialized in a matter of seconds,
\sysname gains this precision 
improvement
with \emph{minimal}
analysis time increase.

Specifically, dynamic execution allows \sysname
to assign unique calling contexts for all
dynamically executed functions invoked
from
multiple call sites.
The precision 
benefits of this calling context separation is significant
if the called function manipulates function pointers
or allocates new objects.
For example, the function \texttt{setup} 
is invoked from three different call sites with three different arguments.
Under a context-insensitive static pointer
analysis algorithm, the points-to sets of 
\texttt{srv->init} and \texttt{req->init} will be merged,
resulting in the loss of precision. On the other hand,
dynamic execution
allows \sysname to
maintain full context sensitivity for the initialization
phase code. We describe the details of our approach 
in Section~\ref{design:buildinitial}.


\subsection{Compounding Precision Improvement}
\label{sec:compounding}
The precision improvements facilitated by the previous two categories
often have a compounding effect during the final static analysis
which improves precision further. This is especially true for 
function pointers, as any imprecision in the call graph can potentially
propagate to the \emph{arguments} of these functions.
For example, when \sysname seeds the initial points-to results, it
can correctly identify that the \texttt{srv->init} 
function pointer can only point to
the function
\texttt{default\_init}. This allows the subsequent static analysis
phase to determine correctly that the indirect function call
\texttt{*(srv->init)()} can only invoke the \texttt{default\_init}
function. 
Thus, the function \texttt{gzip\_init}, which
initializes the pointer \texttt{srv->cb} to \texttt{gzip\_handler},
will be completely excluded from the analysis.
Therefore, only \texttt{default\_handler} will
be added to the points-to set of \texttt{srv->cb}, 
when the function \texttt{default\_init} is statically analyzed.
This compounding precision improvement effect 
is significant in real-world, large, and complex codebases.


%
%

%% file: design.tex
\section{Design}
\label{sec:design}


\sysname provides a custom source code annotation named \texttt{start\_processing},
which developers must use to 
specify the beginning of the processing phase.
The start of the processing phase is typically easily identifiable for 
split-phase applications, and 
is usually denoted by a blocking network system call 
such as \texttt{accept}, or an event loop that uses \texttt{epoll\_wait}. Automated
techniques~\cite{alhanahnah2023slash} can further simplify the identification 
of this transition point.

\begin{figure}[t]
    \centering
    \includegraphics[width=0.969\columnwidth]{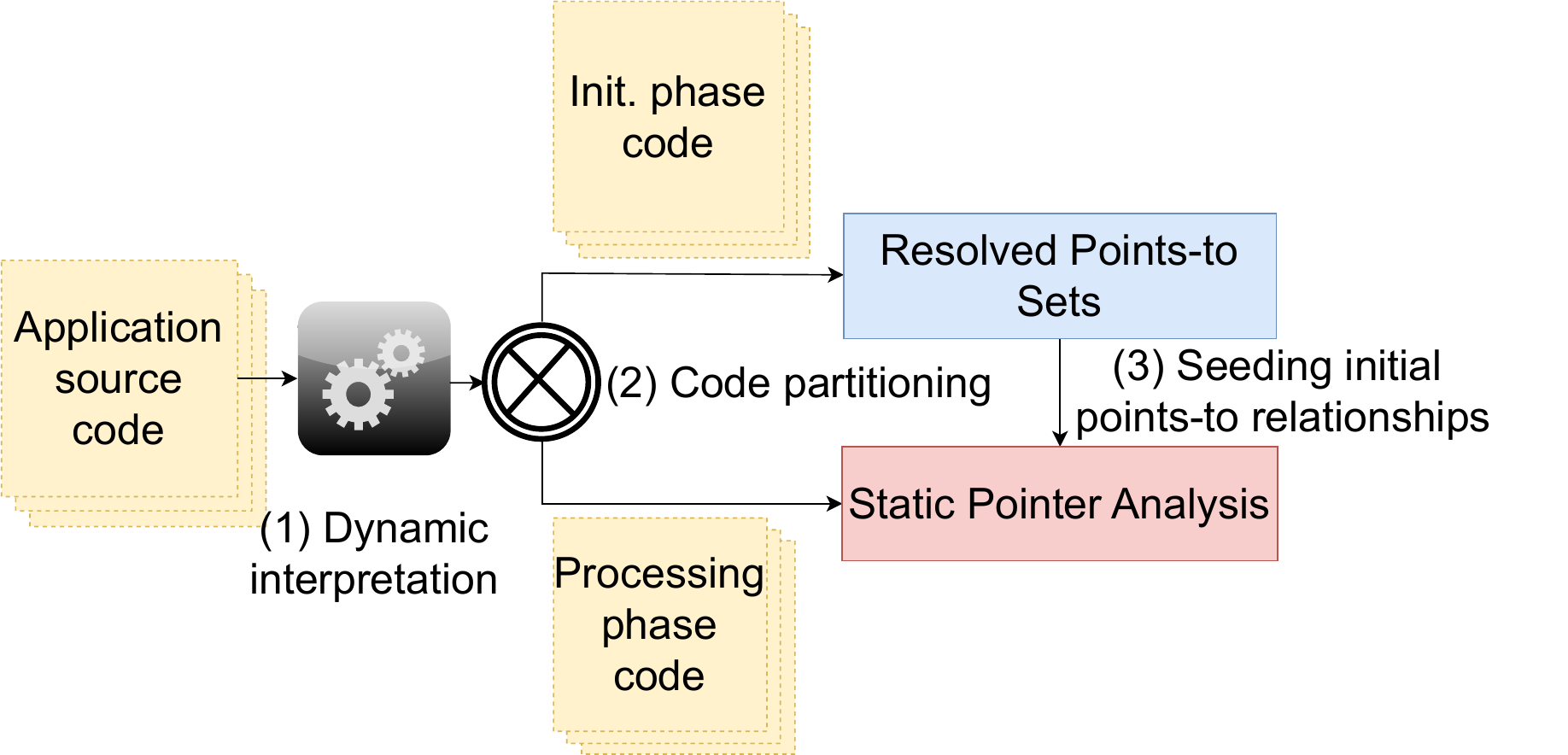}
    \caption{\sysname pipeline stages.}
    \label{fig:dynstablk}
\end{figure}

Provided with this developer input, \sysname begins to dynamically execute
the server application from the initialization phase until it
reaches the annotated code point. Then, it collects the dynamically
observed points-to relationships and switches over to pure static analysis
for the remaining code of the processing phase.
\autoref{fig:dynstablk} provides an overview of the stages in the \sysname pipeline.

\subsection{Dynamic Interpretation}
\label{design:metadata}

The first challenge in seeding dynamic execution results 
into static pointer analysis is the semantic gap
between dynamic execution of binary code and static analysis, 
which is typically performed on source code.
Critical information
such as \texttt{struct} object boundaries
that are essential for the precision of static analysis
are lost when the source code is compiled to the binary format. 
Furthermore, due to various compiler optimizations, it is challenging to precisely map
the assembly instructions in application binaries to source code statements.
To bridge this semantic gap, we perform both
the dynamic execution and the static pointer analysis on an intermediate representation
of the C source code. The dynamic execution is essentially an 
\emph{interpretation} of this intermediate representation, thus allowing
\sysname to gain fine-grained control of this execution.

\sysname's dynamic interpreter begins by loading the program's intermediate 
representation into
its memory. Then, starting from the application's 
\texttt{main} function, it executes
each program instruction, one by one, till it reaches the \texttt{start\_processing}
annotation. Then, it inspects the application's in-memory state and
collects the pointer relationships between the different objects. 
These
dynamically established pointer relationships are seeded to the 
final static analysis phase.
During interpretation, 
the interpreter maintains metadata for all application 
variables, including heap objects,
created and accessed during the initialization phase, including \ding{182} 
a mapping between the variable's name and its 
memory address, \ding{183} a reverse mapping 
that maps a memory address to the variable
name, and \ding{184} a mapping between the variable's name and its type.
\sysname uses this metadata to traverse
the in-memory application objects
to collect the dynamic points-to relationships.

\subsection{Accurate Type Derivation} 
\label{design:typeident}

\begin{figure}[t]
    \centering
    \includegraphics[scale=0.60]{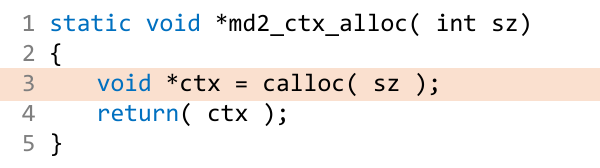}
    \caption{MbedTLS example illustrating difficulty in statically 
    determining heap object types.}
    \label{fig:mbedtlssample}
\end{figure}

Collecting the pointer targets of 
complex in-memory \texttt{struct}
and array objects,
requires the knowledge of the offsets of the pointers embedded
in them. This, in turn, requires the correct type information 
of these objects.
Identifying the type of heap objects is challenging, especially if
the heap 
allocation function provides no type information.
For example, in the case of the simplified code snippet
from the MbedTLS library shown in \autoref{fig:mbedtlssample},
the newly allocated object is stored into a generic 
\texttt{void*} pointer and the \texttt{sizeof} operand is
an opaque integer containing no type information.
Furthermore, the heap allocator function \texttt{md2\_ctx\_alloc} could potentially
be invoked
via an indirect function call, making it even more challenging to track
the object size and type passed. 

%
%
%

To derive the full types of each object
initialized during the initialization phase, \sysname tracks the type-cast 
operations applied on each object
during interpretation. 
Our system hooks
each call to known Libc
memory allocation functions, such as \texttt{malloc}, \texttt{calloc},
and 
\texttt{mmap}, to record both the address
and the size of the allocated object. Then, when a type-cast operation
operating on a heap object
is interpreted, \sysname updates the heap object type according to the
cast type.

\begin{figure}[t]
    \centering
    \includegraphics[scale=0.60]{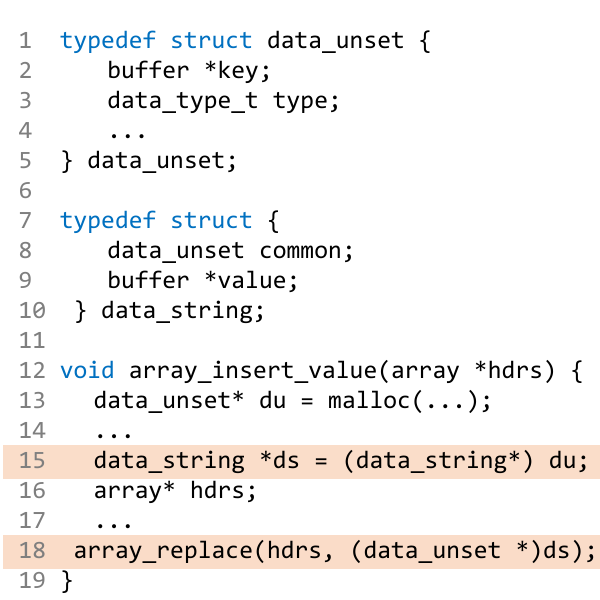}
    \caption{Lighttpd example illustrating a cast \emph{up}
from a more expressive type to a less expressive type. 
}
    \label{dynsta:lightyinherit}
\end{figure}

\para{Type Punning and Type Upcasts.}
In some cases, naively 
updating the type of the heap object for \emph{every} type-cast
operation can result in the loss of type 
information. This is particularly true for applications
that use type punning, as shown in \autoref{dynsta:lightyinherit}.
The \texttt{data\_string}
type
\emph{inherits} from the \texttt{data\_unset} type using
type punning.
In Line 15, the heap object 
pointed to by the pointer \texttt{du},
is first \emph{downcasted} from
\texttt{data\_unset} to the child type
\texttt{data\_string}. 
Then in Line 18, the object is
\emph{upcasted} to
the parent type \texttt{data\_unset}. 
Thus, naively updating the type of each heap object
with the target type for each type-cast operation would result in the association of the type \texttt{data\_unset} with 
the heap object, instead of type 
\texttt{data\_string}, resulting in the loss of the complete
type information for the object.


\sysname uses the notion of \emph{type descriptiveness} 
to handle this potential loss of type information. 
We define
type descriptiveness as the \emph{total} number of fields contained
in a \texttt{struct} type after all nested
\texttt{struct} type fields are expanded. 
Thus, in the Figure~\ref{dynsta:lightyinherit} code snippet, the type
\texttt{data\_string} is more descriptive than 
\texttt{data\_unset}, as it describes the
field \texttt{value} in addition to all of the fields of
\texttt{data\_unset}.
During interpretation, \sysname updates the type 
of a heap object \emph{only if} the object is type-casted to a
\emph{more} descriptive type.
Thus, when the function
\texttt{array\_replace\_value} casts an object of type \texttt{data\_string}
back to \texttt{data\_unset}, which is less descriptive, thus \sysname ignores
this type-cast operation, and retains the type \texttt{data\_string} for the
object, thus maintaining correctness.


%
%
%
%

\para{Heap Arrays.}
\begin{figure}[t]
    \centering
    \includegraphics[scale=0.60]{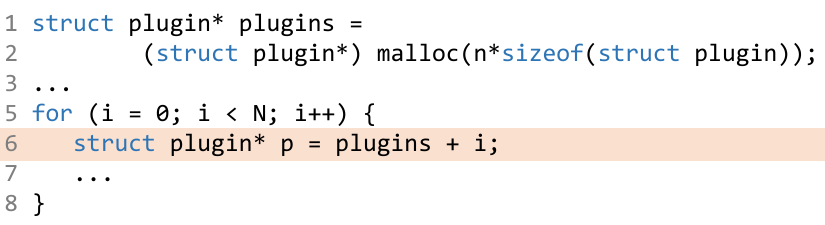}
    \caption{Heap array accesses using pointer arithmetic.}
    \label{fig:heaparray}
\end{figure}
The equivalence of pointer arithmetic and array indexing
in C/C++ further complicates accurate type identification.
In C/C++, a
pointer can be initialized to the first
element of an array and each subsequent element 
can be accessed by using increment operations performed
on that pointer.
In the case of heap arrays, the array
can thus be type-casted to the type of the
\emph{first} element of the array, instead
of being type-casted to the array type.
Thus, simply
extracting the type information from the type-cast operation is not
enough to correctly derive the complete heap array type.
Specifically, the array size information is lost.
The simplified code snippet in ~\autoref{fig:heaparray} shows how 
Lighttpd allocates a variably-sized array of
plugins on the heap and accesses each array element via the pointer \texttt{plugins} 
that is initialized to the first plugin.

To accurately derive the size of heap arrays, the interpreter first
records the size of each heap object
at its allocation site. Then, when the type-cast operation is
interpreted, \sysname \emph{derives} the number of elements in the array
by comparing the size of the allocated heap object
with the size of pointer's \emph{base} type. 
For example, in the 
case of Figure~\ref{fig:heaparray}, the interpreter
divides the size of the heap object, allocated in Line 2 by 
\texttt{sizeof(struct plugin)} to determine
the array size, thereby deriving the complete
array type.

\subsection{Code Partitioning}
\label{design:partition}
When the interpreter encounters \sysname's \texttt{start\_processing}
annotation that marks the transition point,
the application code is partitioned into
functions accessible from the
initialization and the processing phase.  
Performing this partitioning correctly is critical for the soundness
of the analysis.
To achieve sound partitioning, 
\sysname uses an iterative process that relies on the following
criteria for identifying the functions accessible from the processing
phase.


\para{Direct Function Calls.} 
\sysname maintains a list $\mathit{F}$
of functions accessible from the processing phase.
It starts populating this list
by collecting the functions accessible via direct
function calls from the transition point.
To achieve this,
\sysname traverses all basic blocks that are reachable from the
transition point through direct function calls and adds the targets
of all encountered direct function calls to the 
function list $\mathit{F}$.

\para{Initialization Phase Address-Taken Functions.}
The initialization phase might potentially store the
addresses of certain functions to function pointers.
To handle such address-taken functions
\sysname scans the application's global, stack, and heap
objects for any values that correspond to
function addresses and adds them to the accessible list
of functions $\mathit{F}$.
These are the functions whose addresses are taken and
stored during the initialization phase and can be 
invoked via indirect call sites during
the processing phase. Function pointers contained
in global variables require special handling. We
consider a global variable (and any functions it refers to) to be
\emph{accessible} only if it either (a) is directly accessed
from an accessible function, or (b) another accessible stack, heap, 
or global variable holds a reference to it. In other words, if 
the global variable cannot be accessed \emph{at all} from the
processing phase, we exclude any function pointers it contains from
the processing phase code.

\para{Processing Phase Address-Taken Functions.} 
Finally, the processing phase might store and update
function pointers. To 
handle these cases,
\sysname scans the program statements of all functions in the
accessible list $\mathit{F}$, for any instructions
that store the address of any \emph{new} function to a
function pointer. All such functions are added
to the list $\mathit{F}$.

Putting these strategies together, we develop an
iterative algorithm that recursively
collects all functions that might be accessed
from the processing phase
till a \emph{fixed point} is reached and no new 
accessible function is discovered.
This ensures that our algorithm can 
soundly compute all functions 
that remain in scope at the end of the 
initialization phase, with respect
to the given runtime configuration.
The complete algorithm for this process is described in
Appendix~\ref{app:code-part}.


\subsection{Seeding Initial Points-to Relationships}
\label{design:buildinitial}

After identifying which functions can be invoked from the processing phase,
\sysname proceeds to build the points-to graph of all pointers and objects
initialized during the initialization phase and \emph{seeds}
the static analysis phase with
this graph. The goal of this step is to
\emph{pre-solve}, with full precision,
the pointers and objects that are created till the
transition point. 
Note that the points-to relationships established in this phase 
are not final---the subsequent static
analysis phase processes and potentially updates 
these relationships.

To build the initialization phase points-to graph, \sysname iterates over every
pointer in application memory and reads the value stored in its 
memory. Then, using the interpreter's 
internal metadata maps and collected type information, it resolves
the points-to relationships by comparing the pointer values
to the addresses of in-memory objects. These dynamic points-to 
relationships are then seeded to the final static pointer analysis
stage.
%

\para{Context Sensitive Heap Seeding.}
\label{sec:contextsensitiveheapseeding}
Seeding the dynamic points-to relationships to the static analysis phase
requires mapping the in-memory addresses to IR program
statements
that define or create objects. 
This process
is straightforward in the case of global variables
and stack objects as the object address has a 1-to-1 mapping
with a program statement that defines the 
corresponding object. But in the case of heap objects, the same program
statement (e.g., a call to the \texttt{malloc} function) 
can be invoked
multiple times from different \emph{calling contexts},
resulting in the
creation of multiple heap objects. Each of these objects
can have their own unique
points-to relationships. 
Mapping \emph{all} of these \emph{unique} objects
to the same heap allocation program statement would 
prevent the static analysis from
distinguishing between these unique heap objects, resulting
in loss of precision.

\sysname uses \emph{cloning} to preserve the precision of the
heap objects created during the initialization phase. During interpretation,
the interpreter maintains the
mapping between each memory allocation program
statement (\texttt{malloc, mmap, calloc},
etc.) and the heap objects it allocates at runtime. For each heap-allocated
object at a given call site, \sysname \emph{clones} the 
memory allocation program statement, 
and ensures that a unique heap allocation statement corresponds to each 
heap object created during dynamic execution.

\begin{figure}[t]
    \centering
    \includegraphics[scale=0.44]{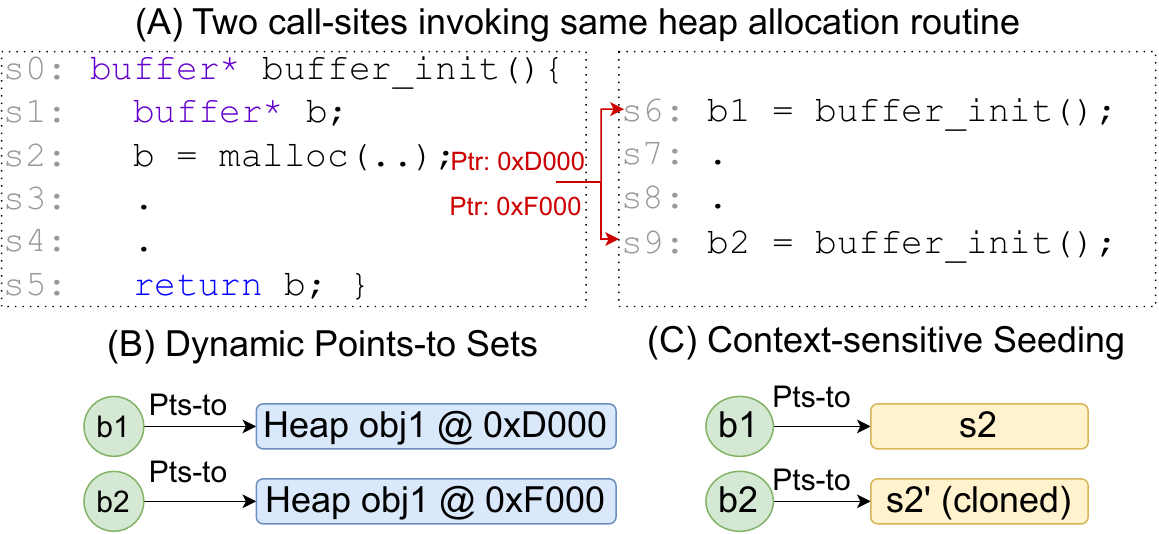}
    \caption{Context-sensitive seeding of heap objects' pointer relationships.} 
    \label{fig:context-sensitive-heap}
\end{figure}

%
%
Figure~\ref{fig:context-sensitive-heap} shows a simplified 
illustration from the Lighttpd codebase, where the \texttt{malloc}
invocation in statement \texttt{s2} returns two different heap
objects 
with addresses \texttt{0xD000} and \texttt{0xF000}, 
to the statements \texttt{s6} and \texttt{s9}, respectively. 
Cloning the statement \texttt{s2} allows \sysname to
separate the calling contexts and 
precisely seed
the points-to relationships for pointers \texttt{b1} and \texttt{b2}.

\para{Array Element Seeding.}
The static pointer analysis framework used by \sysname is
array-index insensitive, and 
does not distinguish between different elements of an array.
However, the points-to relationships captured by the interpretation
phase distinguish between every array element. 
Therefore, when seeding the pointer relationships
of array elements, we must bridge
this gap between the array-index sensitive 
dynamic interpretation and 
array-index insensitive static pointer analysis.

\begin{figure}[t]
    \centering
    \includegraphics[scale=0.35]{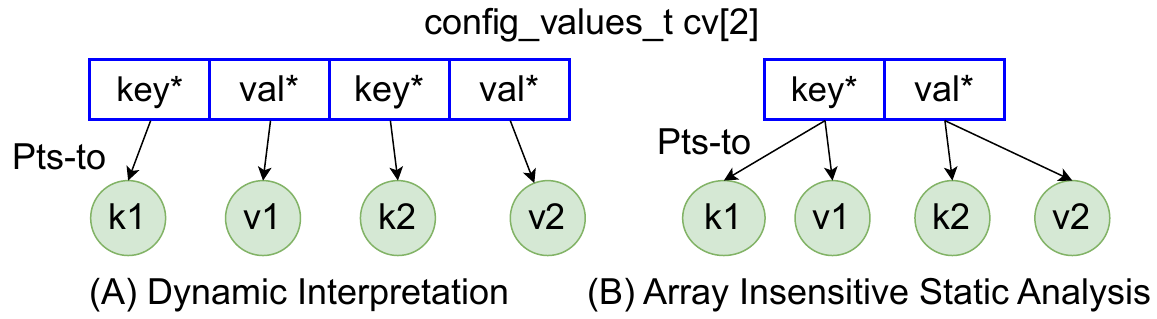}
    \caption{Array-index insensitive pointer
    relationship seeding.} 
    \label{fig:arrayindex}
\end{figure}

We bridge this gap by \emph{collapsing} all array
elements 
during the seeding phase. Consider the simplified code snippet from 
Lighttpd in
Figure~\ref{fig:arrayindex}. The object \texttt{cv} is an array of type
\texttt{config\_values\_t}.
This array is
initialized during server startup, and therefore at the end of the
initialization phase 
we
collapse the corresponding fields of each array element into a single
field as shown. This results in a loss of precision but maintains soundness,
and is unavoidable without using a more expensive array-index sensitive static analysis.




Once the processing phase code is identified and the dynamic points-to 
results are seeded, the static analysis phase 
builds the final points-to sets for all the pointers in the 
application iteratively. \sysname uses
a standard field-sensitive, context and array-index insensitive 
Andersen's-style pointer analysis~\cite{andersen, svf}.

\subsection{Alternate Runtime Configurations Options}
 
Given the substantial increase in the use of containers for deploying 
applications~\cite{cncf-survey}, obtaining commonly utilized runtime
settings has become feasible. Furthermore, due to the increasing
complexity of the runtime settings of modern server applications~\cite{xu2015hey},
developers often rely on the pre-configured server applications that ship with popular Docker
images~\cite{zend-image,dockerhub-wordpress,dockerhub-wordpress-bitnami} and launch their web applications
on top of fully configured web servers such as Nginx.
\sysname can harden these web servers using these fixed runtime configuration options.

To support multiple runtime configurations, we extend \sysname with a lightweight runtime mechanism 
that allows the binary to adapt its CFI instrumentation based on the runtime configuration options with 
which it is launched. 
The binary is shipped with the interpreter and, on launch, checks whether it has already been executed under 
the current configuration. If not, it invokes the interpreter to execute the initialization phase, 
perform pointer analysis, and generate the final CFI results. These results are written to a \texttt{cfi.map} file 
and used to patch the binary before transferring control. After update, the \texttt{cfi.map} is
changed to read-only. For previously seen configurations, the corresponding \texttt{cfi.map} is loaded at launch and the 
binary is patched directly. This approach enables \sysname to apply its precision benefits across diverse runtime environments without sacrificing soundness.

%% file: impl.tex
\section{Implementation}
\label{sec:implementation}

We developed \sysname using the LLVM 12~\cite{llvm} toolchain 
and the popular
SVF~\cite{svf} static analysis library. First, we generate the
Intermediate Representation bitcode for the application using
Link Time Optimization (LTO). The LLVM toolchain provides the 
\texttt{lli} interpreter that can interpret LLVM Intermediate Representation, which we use for dynamic interpretation.
We enhance
the interpreter to record the address and type metadata for the heap objects, by instrumenting
\texttt{lli}'s handling of 
the LLVM \texttt{BitCastInst} 
and \texttt{CastInst}
instructions. 
While \texttt{lli} supports most common
LLVM IR instructions, we had to add support for the
\texttt{atomicrmw} and \texttt{cmpxchg} instructions
to fully support the applications in our
test suite.
We use the default field-sensitive and context-insensitive
Andersen's style pointer analysis algorithm from the
popular SVF~\cite{svf} library for static pointer analysis.
Additional 
details including the handling of 
multithreaded applications is provided in \refappendix{app:imp}.

%% file: usecases.tex
\section{Use Cases}

We apply \sysname to three use cases---
CFI, software debloating,
and system call filtering.
In each of these cases, we use \sysname to 
resolve the indirect call sites and generate the
call graph for the application's processing phase.
Then, using this call graph, we either implement
the CFI checks, remove unneeded functions, 
or filter the unneeded system calls, 
depending on the use case.


\para{Control Flow Integrity.}
We implement a forward-edge CFI mechanism based on the call graph
generated by \sysname. This call graph provides the list of 
permitted targets
for all
indirect call sites invoked during the processing phase. 
Using these lists of permitted targets, we instrument 
each indirect call site to check if the target observed
at runtime is permitted or not. 

Special handling is required if the same indirect call site 
is invoked from both
the initialization and the processing phase with different
permitted targets.
In these cases, we clone and specialize 
the functions containing such indirect call sites into two 
versions, one for
each phase respectively. Then, we apply the corresponding
CFI filters for each clone.
This
allows us to enforce different CFI profiles for both the 
specialized versions correctly.

\para{Software Debloating.}
\label{usecase:debloat}
Software debloating aims to reduce 
an application's attack surface by identifying and
removing unneeded application code.
Generating a sound and precise call graph for the application and its libraries
is essential for performing software debloating statically. An unsound call
graph can break the program, and an imprecise call graph limits the security
guarantees.
Therefore, we use \sysname to generate a more
precise and sound call graph.

Previous works~\cite{piecewise,agadakos2019nibbler} that debloat code statically only
remove code from the application's dependent libraries.
These techniques
consider the entire application a monolithic entity and assume
\emph{all} its code is
required, and only specialize the dependent libraries according
to the application's requirements.
However, by using the processing phase call graph generated by
\sysname we can also debloat application functions that are unneeded under
the given initial configuration, once the application has
finished initialization.


\para{System Call Filtering.}
Filtering unused system calls is
an effective mechanism for reducing the application's attack 
surface~\cite{confineraid20, temporal, sysfilter, syspart}
Similarly to software debloating, system call filtering
also requires the precise call graph of the
application and its libraries to identify the unreachable
system calls.
Temporal Specialization~\cite{temporal} is a technique that uses
static analysis to perform system call filtering.
Temporal Specialization also focuses on two-phase applications,
and uses static analysis to partition the application
code into functions accessible from the initialization 
and processing phases.
Therefore, this technique's security guarantees depend on the precision
of the initialization and processing phase call graphs.

To improve precision, Temporal Specialization 
augments pointer
analysis and call graph construction with argument type  matching, i.e., 
it removes all targets
in the points-to set of an indirect call site whose formal 
argument types
do not match the actual call site argument types. 
However, as discussed in Section \ref{bg:typepunning}, the 
pointer
types derived statically are often unreliable. 
Therefore, performing this type-based 
filtering technique can potentially lead to unsoundness.

We replaced the static call graph generation component of
Temporal Specialization with \sysname. The rest of the 
toolchain 
proceeds unmodified and extracts the system call profiles for each 
stage of the application. Moreover, as discussed in 
Section~\ref{design:typeident}, because \sysname only performs
conservative and sound type-based 
optimizations, the generated call graph is 
guaranteed to be sound.

%% file: eval.tex
\section{Experimental Evaluation}
\label{sec:eval}

We evaluate our
system with eight popular server and desktop 
split-phase applications.
Across all of these applications, we verified that
the
points-to relationships established at the end of the 
initialization phase indeed remained
constant across multiple runs. We also verified that
the applications' functionality was not affected after
hardening them using \sysname's points-to results, demonstrating the
soundness of our approach.

\subsection{Applications}
Table~\ref{eval:apps} presents the applications we used to evaluate our
system, along with the transition function that marks the start of the processing
phase. Nginx~\cite{nginx}, Lighttpd~\cite{lighttpd}, 
and Monkey~\cite{monkey} are web servers
supporting pluggable modules. Of these, Monkey is multi-threaded
while Nginx and Lighttpd use asynchronous I/O using 
\texttt{epoll} for concurrency. Memcached~\cite{memcached}
is an 
event-driven key-value store that uses 
LibEvent~\cite{libevent}. MbedTLS~\cite{mbedtls} is an
SSL/TLS library, and we use the sample \texttt{ssl\_server2}
application for our evaluation. Wget~\cite{wget} and 
Curl~\cite{curl} are command-line utilities for
downloading web content. Both programs provide a
variety of command-line options
that determine how the URL is 
parsed and handled.
Each of these applications
are executed with their default initial configuration.
Four out of the eight applications in our evaluation set use
type-punning, while all of them use \texttt{void*} pointers
to point to complex \texttt{struct} type objects. 
Therefore, 
using a type-based precision improvement technique would potentially
lead to unsoundness for these applications. 

\renewcommand{\arraystretch}{1.05}
\begin{table}[t]
    \centering
    \caption{Applications used for evaluation and whether they use type-punning and
\texttt{void*} pointers. (\cmark: uses
pattern, \xmark: does not use pattern) 
}
    \label{eval:apps}
    \scalebox{0.90}{
        \begin{tabular}{P{1.5cm}P{4.0cm}C{1.25cm}C{1.25cm}}
        \toprule
\bfseries Application & \bfseries Transition Point & \bfseries Type-punning &
\bfseries \texttt{void*} pointers \\
        \midrule
            Nginx         		&
\small{\texttt{ngx\_worker\_process\_cycle}} 	    & \xmark		&
\cmark	\\
            Lighttpd         		&  	 \small{\texttt{server\_main\_loop}}
&   \cmark 			& \cmark \\
            MbedTLS          & 	 \small{\texttt{mbedtls\_net\_accept}}
& 	\xmark			& \cmark  \\
            Monkey      		&  	 \small{\texttt{mk\_server\_loop}}
& \cmark	& \cmark \\
            Ssh-agent               &  	 \small{\texttt{prepare\_poll}}
& \cmark	& \cmark \\
			Memcached               & 	 \small{\texttt{event\_base\_loop}}
& \cmark	& \cmark \\
            Wget       				& 	\small{Basic block in \texttt{main}}
& \xmark	& \cmark \\
            Curl               		&	\small{Basic block in \texttt{operate}}
& \xmark	& \cmark \\
\bottomrule
\end{tabular}
}
\end{table}

\renewcommand{\arraystretch}{1.05}
\begin{table}[]
\caption{Points-to set sizes for whole program static pointer
analysis using SVF vs. \sysname.}
\label{eval:ptssetsize}
    \scalebox{0.90}{
        \begin{tabular}{P{1.9cm}E{1.5cm}E{1.3cm}E{1.5cm}E{1.3cm}}
    \toprule
                     & \multicolumn{2}{r}{\textbf{Max Pts-to Set Size}}  &
\multicolumn{2}{r}{\textbf{Avg Pts-to Set Size}} \\
\bfseries Application & \bfseries SVF & \bfseries \sysname & \bfseries SVF & \bfseries \sysname  \\
    \midrule
Nginx	      & 1837	& 1465	  & 372.46	   & 316.24 \\
Lighttpd	  & 1433	& 911	  & 70.14	      & 11.86 \\ 
Mbedtls	      & 244 	& 244	  & 31.21	      & 28.63 \\
Monkey	      & 795	    & 247	  & 198.97	      & 48.11 \\
Ssh-agent	  & 182    	& 105	  & 1.46	      & 1.26 \\ 
Memcached	  & 757	    & 298	  & 38.43	      & 16.24 \\
Wget	      & 489    & 460      & 17.19	     & 14.87 \\
Curl	      & 2059	& 1782	 & 212.98	& 147.50 \\
\bottomrule
\end{tabular}
}
\end{table}

\subsection{Pointer Analysis Statistics}
Pointer
analysis imprecision results in the inclusion of spurious objects
in the points-to sets, i.e.,
the analysis derives pointer relationships 
that are not possible during actual runtime execution.
Therefore,
the size of the points-to sets is an indication of the level of imprecision of
the pointer analysis. Table~\ref{eval:ptssetsize} shows the average and
maximum points-to set size for \sysname 
and for the standard field-sensitive
and context-insensitive pointer
analysis provided by the popular SVF framework~\cite{svf}.
\sysname significantly
reduces both the average and maximum points-to set sizes. 
For all applications except Curl and MbedTLS, 
\sysname shows a significant precision improvement. We discuss
the reasons for this behavior in \autoref{eval:cfi}.
\refappendix{app:analysistime} evaluates the analysis time
for the baseline fully static pointer analysis and \sysname.
Across all applications, \sysname takes less time
than the baseline fully static pointer analysis.

\begin{figure*}[t]
    \centering

	\subfloat[][Average EC Size]{
    \label{fig:avgcfi}
    \includegraphics[width=0.48\linewidth]{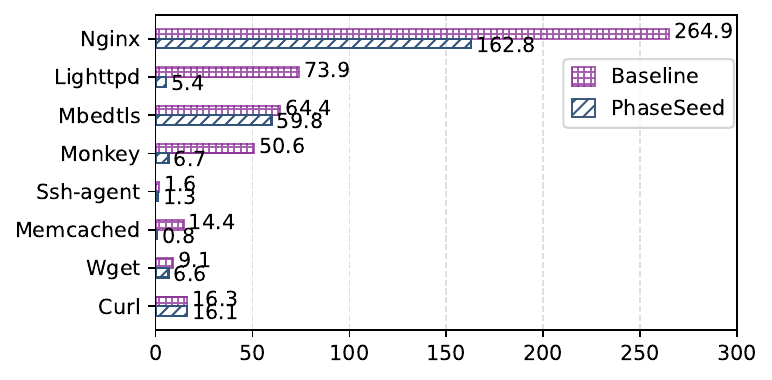}
    }
      \hspace{0.1em}
    \subfloat[][Maximum EC Size]{
    \label{fig:maxcfi}
    \includegraphics[width=0.48\linewidth]{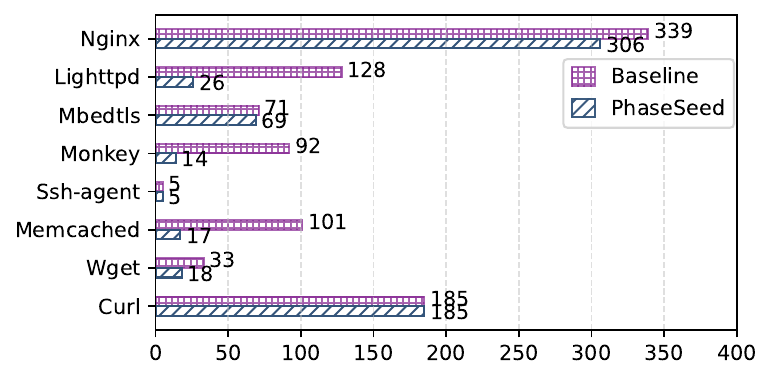}
    }
    \caption{Average and Maximum EC (Equivalence Class) size for CFI.}
    \label{fig:cfi}
\end{figure*}

\begin{figure*}
    \centering
	\subfloat[][Program Function Count]{
    \label{fig:fncount}
    \includegraphics[width=0.48\linewidth]{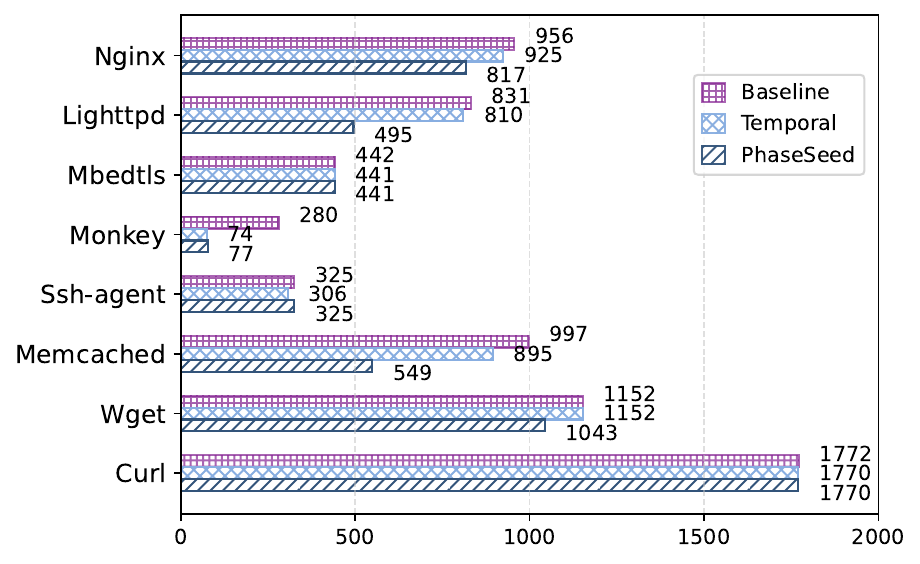}
    }
      \hspace{0.1em}
    \subfloat[][Program IR Instruction Count]{
    \label{fig:ircount}
    \includegraphics[width=0.48\linewidth]{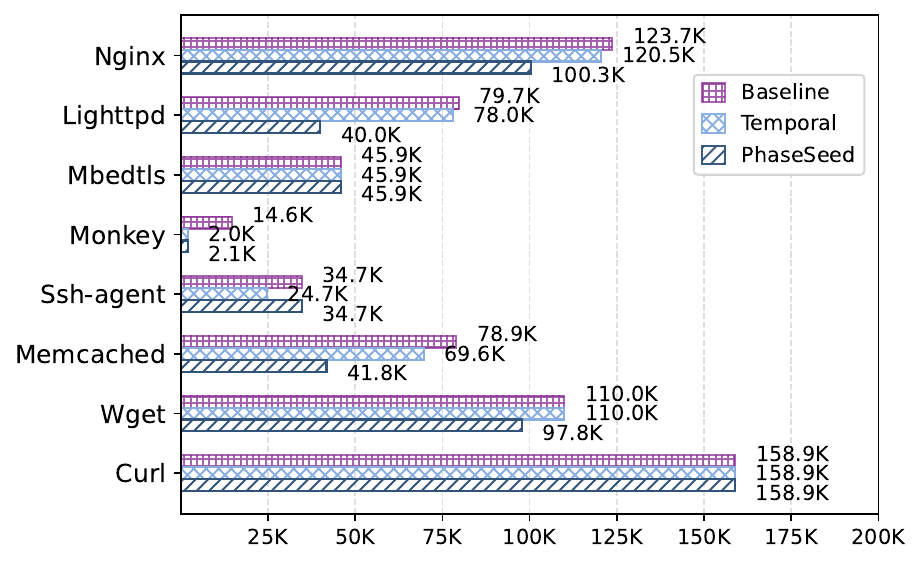}
    }
    \caption{Number of functions and IR instructions debloated.
    }
    \label{fig:debloating}
\end{figure*}

\subsection{Control Flow Integrity}
\label{eval:cfi}

\autoref{fig:cfi} reports the maximum and average number of targets for each
indirect call site using the naive SVF-based approach and our
technique. The targets for each indirect call site
are considered an \emph{Equivalence Class (EC)}.
All server applications
except Curl and MbedTLS, 
show an improvement in the precision of
the resolved targets ranging from 43\% to 92.6\%. 
The web servers Nginx and Lighttpd,
support configurable 
\emph{modules} which encapsulate custom functionalities
that can be enabled by the initialization configuration.
These modules
are created and stored on the heap during the 
initialization phase, 
and thus they
benefit from the fully precise dynamic analysis
and context-sensitive seeding of heap
objects, as discussed in Section~\ref{design:buildinitial}.
These are examples that clearly show the benefit of dynamic seeded pointer
analysis. Note that in the case of Memcached, the average EC size
is less than 1 because some indirect call sites in the Libevent 
library have no targets because those Libevent features are
not used by Memcached.

In the case of MbedTLS, however,
the precision improvement is only marginal. 
This is because, in the case of MbedTLS, the initialization phase
simply loads and configures the private keys and does not set up any function
pointers. The main use of function pointers 
in the MbedTLS codebase occurs
in the encryption/decryption routines 
of the different protocols. These
protocols however are stored in different global 
\emph{protocol objects} and are
invoked depending on the type of client request received,
which is resolved during the processing phase.
Therefore, all global
protocol objects remain \emph{in scope} during the 
processing phase, and \sysname is forced to conclude that all
encryption/decryption protocol functions are valid at every indirect call site.  
Curl also performs limited operations during initialization,
mainly parsing
configuration settings without allocating 
any objects that persist into the processing phase.
Therefore, \sysname provides lower precision improvements for
this application.
\refappendix{app:ablation} provides an ablation study 
to determine the precision
improvements provided by each \sysname  
component.

\subsection{Software Debloating}
Similar to prior works on software debloating, we report the 
amount of code that remains
accessible to assess \sysname's attack surface reduction benefit. 
Since previous works~\cite{piecewise,agadakos2019nibbler} on code debloating
through static analysis
do not improve the precision of the
application's call graph, we cannot use them to perform a proper comparison. 
Instead, we use the call graphs generated by Temporal
Specialization~\cite{temporal} as our point of comparison. 
Figure~\ref{fig:debloating} compares
the number of accessible functions and LLVM IR
instructions for the 8 applications in our
dataset
using the call graph generated by the baseline SVF, Temporal Specialization,
and \sysname.

\sysname provides a significant code reduction benefit for Lighttpd, Memcached, and
the Monkey server, reducing the
number of accessible functions from 947 to 573 for Lighttpd, 1000 to 549 for
Memcached, and 355 to 83 for the Monkey server. 
The number of accessible IR instructions after
applying \sysname is reduced by 48\%, 47\%, and 86\% for Lighttpd, Memcached, and the
Monkey server, respectively.
\sysname also reduces the accessible functions and instructions for Nginx and Wget.
The code reduction for MbedTLS, Ssh-agent, and Curl is 
limited, which occurs
for the same reasons described in Section ~\ref{eval:cfi}.


\subsection{System Call Filtering}
\label{sec:syscall}
We evaluated the number of system calls that can be 
filtered by \sysname
compared with the original Temporal 
Specialization~\cite{temporal} toolchain.
As shown in Table~\ref{eval:syscall},
\sysname filters additional system calls for Nginx and Lighttpd
compared to Temporal Specialization~\cite{temporal},
offering a considerable attack surface reduction.
These include
system calls, 
such as \texttt{bind} and \texttt{select} for Lighttpd, which have been proven to be
security-critical by previous 
work~\cite{temporal}, as discussed in \refappendix{app:security-critical}.
Note that Lighttpd supports multiple mechanisms for processing
incoming requests, including \texttt{epoll}, \texttt{select},
and \texttt{poll}. The default initial configuration for Lighttpd
enables only \texttt{epoll},
which allows \sysname to filter the 
\texttt{select} and \texttt{poll} system
calls.
For the other applications, \sysname 
removes the same number of 
system calls as Temporal Specialization. 
Moreover, \sysname's call graph analysis avoids the 
soundness issues of Temporal Specialization’s type-based matching.

\renewcommand{\arraystretch}{1.2}
\begin{table}[t]
    \centering
    \caption{Additional system calls filtered by \sysname compared to Temporal Specialization.}
    \label{eval:syscall}
    \scalebox{0.90}{
        \begin{tabular}{P{1.8cm}P{6.5cm}}
        \toprule
\bfseries Application & \bfseries Additional System Calls Filtered  \\
        \midrule
            Nginx         		& \texttt{sched\_yield}, \texttt{pwritev}, \texttt{rename}, \texttt{utimes}\\
            Lighttpd         		&  \texttt{bind}, \texttt{listen}, 
                                    \texttt{select}, \texttt{pipe2}, \texttt{poll} \\
\bottomrule
\end{tabular}
}
\end{table}

%% file: discussion.tex
\section{Discussion}
\label{sec:discussion}

\parasmall{Dynamic vs. Static Boundary.}
\sysname assumes
that the transition point splits the program into two
phases. While this separation is adequate to illustrate
the benefits of \sysname, further precision improvements can potentially 
be obtained
by the addition of deeper phases. 
This is especially true for applications such as Curl, 
which continue parsing configuration options
past \sysname's selected transition point. 
In the case of such applications, moving the
transition point deeper into the application code would improve
the precision further. 
Placing the transition point beyond the start of the serving
phase would cause unsoundness, 
but \sysname can detect this 
because in that case, the developer-placed transition
point would not be reached during the dynamic
interpretation as the application
waits for a user request. Thus, \sysname will flag this
error and protect against potential unsoundness.



\para{Other Use Cases.}
\sysname is a generic technique that improves pointer analysis precision and
can be applied to any use case that requires pointer analysis. 
DSR~\cite{dsr} and DynPTA~\cite{palit2021dynpta}
use static pointer analysis to obtain
the superset of all pointers that might point to sensitive application 
data to randomize or encrypt it. 
DPP~\cite{dpp} uses pointer analysis to
identify security-critical application data. Similarly, fuzzing
techniques~\cite{hawkeye,muzz} use pointer analysis to identify and 
prioritize interesting mutations. 
\sysname can be used to improve the precision of these
mechanisms. 
 
%

%% file: related.tex
\section{Related Work}

\parasmall{Pointer Analysis Precision Improvement.}
Traditional techniques for improving precision 
require 
adding \emph{more} program information to the pointer analysis. 
Previous works~\cite{cloningbased, efficientcontextsen}
improve context sensitivity using cloning-based
and summarization-based approaches.
Hardekopf and Lin~\cite{hardekopf2009semi} proposed
using Binary Decision Diagrams (BDDs) to implement flow and 
context-sensitivity.
Pearce 
et al.~\cite{pearce2007efficient} presented 
a field-sensitive pointer analysis. Lei et al.~\cite{lei2019fast} improved
the precision field derivations by using object type information.
Hasti et al.~\cite{hasti1998using} iteratively converted memory to SSA form
to gradually add flow-sensitivity to their points-to results.
Li et al.~\cite{li2023hybrid} propose a hybrid approach that combines 
data flow analysis and type-based analysis
to identify writes to global variables in the Linux kernel.
Integrating \sysname 
with these techniques can further improve their
precision.
Recent works~\cite{Tongpex2019,shard,lu2023practical,cai2024unleashing} leverage a
type-based analysis to generate call graphs, including
for large codebases such as the Linux kernel~\cite{Tongpex2019}.
These approaches suffer from the limitations discussed
in Section~\ref{bg:typepunning}. 
Kallgraph~\cite{kallgraph} diagnoses soundness and completeness gaps in 
type-based call graph construction and remedies them by combining type-based analysis 
with demand-driven pointer analysis. PhaseSeed instead provides a general refinement 
framework for pointer analysis, of which call graph generation is just one application.

%
%
%
%
%
%


Past-Sensitive pointer 
analysis~\cite{pastsen} integrates pointer analysis with symbolic execution to improve its precision.
Hybrid pruning~\cite{hybrid} applies 
dynamic profiling to derive pointer relationships, but does not
provide any soundness guarantees.
Optimistic Hybrid Analysis~\cite{oha} uses predicated
static analysis to accelerate dynamic analysis. Smaragdakis et al.~\cite{heapsdont} use dynamic heap snapshots to improve
the soundness of static analysis to account for dynamic loading and
cross-language libraries. In contrast with these techniques, \sysname
presents a sound technique for seeding the results of dynamic execution
to improve pointer analysis precision.

\para{Control Flow Integrity.}
CFI was originally introduced by Abadi et al.~\cite{cfi2005}
to protect against control flow hijacking attacks. 
Since then, 
many works have discussed its 
shortcomings~\cite{carlini2015control,evans2015control,conti2015losing} and 
proposed methods to
improve its precision and 
effectiveness~\cite{criswell2014kcofi,van2015practical,christoulakis2016hcfi,khandaker2019origin,niu2014modular,hu2018enforcing,gu2017pt,ding2017efficient,van2016tough}.
More recent
works~\cite{gu2017pt,ding2017efficient,hu2018enforcing,van2015practical} 
leverage hardware features (e.g., Intel PT) to obtain runtime information and
reduce the valid targets for each indirect call site. 
Such techniques are orthogonal to \sysname. 
TypeArmor~\cite{van2016tough} uses type-based matching, and
OSCFI~\cite{khandaker2019origin} leverages data flow analysis to 
reduce the number of targets of indirect call sites. 
However, previous
work~\cite{li2020finding} shows that these techniques are unsound
and result in compatibility issues making their usage impractical.

\para{Software Debloating.}
Various works have focused on attack surface reduction 
using static
analysis~\cite{temporal,davidsson2019esorics,sysfilter,piecewise,saffire,c2c,syspart}, dynamic
analysis~\cite{ghaffarinia2019binary,qian2019razor,codaspy-codespec}, 
or hybrid approaches~\cite{confineraid20,lightblue-sec21,canella2020automating,qian2020slimium,porter2020blankit}.
Callgraph construction underpins many of the recent techniques~\cite{piecewise, agadakos2019nibbler, sysfilter, jelesnianski2023protect, canella2020automating}.
%
Saffire~\cite{saffire} performs argument-level specialization 
and requires an accurate call graph to resolve the argument flows
via indirect calls.
Configuration-to-Code (C2C)~\cite{c2c} maps runtime
settings to the application's code through static analysis
and filters unnecessary system calls using this mapping.
\sysname can generate a more precise call graph improving these works' 
effectiveness.

%% file: conclusion.tex
\section{Conclusion}
\label{sec:conclusion}
We presented \sysname, a novel technique for
precise call graph construction for split-phase applications.
We applied \sysname to control 
flow integrity (CFI), software debloating, and system call filtering,
and showed that \sysname provides up to 92.6\% precision improvement 
compared to state-of-the-art static
call graph construction techniques.

%% file: appendix.tex
\appendices
\label{sec:appendix}
\section{Static Pointer Analysis Constraints}
\label{app:cons}
Static pointer analysis consists of two stages. In the first stage, 
all pointer-related operations are converted to constraints. The second
stage solves these constraints according to the constraint resolution
rules. This section describes the constraints and the 
constraint resolution rules for Andersen's style~\cite{andersen} pointer
analysis.

Four types of pointer operations result in the creation of
constraints. These are described below. Note that $p$ and $q$ in these examples can be single-level (\texttt{int
*p}) or
multi-level (\texttt{int **p}) pointers.

\begin{enumerate}
    \item $p \coloneqq \&x$ (\emph{Address-of})
    \item $p \coloneqq q$ (\emph{Copy})
    \item $p \coloneqq *q$ (\emph{Dereference})
    \item $*p \coloneqq q$ (\emph{Assign})
\end{enumerate}

The points-to set for a pointer $p$ is depicted by $pts(p)$. 
The constraint resolution rules for Andersen's
inclusion-style analysis are as follows:

\begin{enumerate}
    \item $p \coloneqq \&x \Rightarrow x \in pts(p)$
    \item $p \coloneqq q \Rightarrow pts(p) \supseteq pts(q)$
    \item $p \coloneqq *q \Rightarrow pts(p) \supseteq pts(pts(q))$
    \item $*p \coloneqq q \Rightarrow pts(pts(p)) \supseteq pts(q)$
\end{enumerate}

The pointer analysis algorithm iteratively applies these rules
to update the points-to set of each pointer in the application, until
the \emph{fixed point} is reached and no new pointer 
relationships are observed.

\section{Code Partitioning Algorithm}
\label{app:code-part}
The complete algorithm for the code partitioning stage
described in Section~\ref{design:partition} is shown in 
Listing~\ref{dynsta:algo}. The algorithm iteratively
accumulates all functions potentially reachable
from the transition point, until a fixed point is reached
and no new functions are discovered.

\begin{algorithm}[]
\SetAlgoLined
    \SetKwInOut{KwIn}{Input}
    \SetKwInOut{KwOut}{Output}
\SetKwRepeat{Do}{do}{while}
\KwIn{$HeapObjects$ = In scope heap objects at the transition point \newline
$StackObjects$ = In scope stack objects at the transition point \newline
$GlobalVariables$ = All global variables in the application} 
\KwResult{$F$ = set of functions accessible from the processing phase. }
    $F$ = $\emptyset$ \;
    $GlobalToFnMap$ = $\emptyset$ \;
    $AccessibleGVs$ = $\emptyset$ \;
    \ForEach{$H_{obj}$ $\in$ $HeapObjects$} {
        Inspect $H_{obj}$ to find all contained function references $F_h$\;
        $F$ = $F$ $\cup$ $F_h$\;
        Inspect $H_{obj}$ to find all global variable
        references $G_h$\;
        $AccessibleGVs$ = $AccessibleGVs$ $\cup$ $G_h$\;
    }
    \ForEach{$Stk_{obj}$ $\in$ $StackObjects$} {
        Inspect $Stk_{obj}$ to find all function references $F_{stk}$\;
        $F$ = $F$ $\cup$ $F_{stk}$\;
        Inspect $Stk_{obj}$ to find all global variable
        references $G_{stk}$\;
        $AccessibleGVs$ = $AccessibleGVs$ $\cup$ $G_{stk}$\;
    }
    \ForEach{$GVar$ in $AccessibleGVs$} {
        Inspect $GVar$ to find all function references $F_g$\;  
        $F$ = $F$ $\cup$ $F_g$\;
    }
    \ForEach{G $\in$ $GlobalVariables$} {
        Inspect $G$ to find all function references $F_g$\;
        $GlobalToFnMap[G]$ = $GlobalToFnMap[G]$ $\cup$ $F_g$\;
    }
    $F_{new}$ = $F$\;
    \Do{$F_{new}$ $\neq$ $\emptyset$ } {
        $F$ = $F$ $\cup$ $F_{new}$\;
        $F_{new}$ = $\emptyset$\;
        \ForEach{$func$ $\in$ $F$} {
            \ForEach{$insn$ $\in$ func}{
                \If{$insn$ is a direct call to function $T$} {
                    $F_{new}$ = $F_{new}$ $\cup$ $T$\;
                }
                \If{$insn$ takes address of function $T$} {
                    $F_{new}$ = $F_{new}$ $\cup$ $F$\;
                }
                \If{$insn$ accesses the global variable $G$} {
                    $F_{new}$ = $F_{new}$ $\cup$ $GlobalToFnMap[G]$\;
                }
            }
        }
    }
    $F$ = $F$ $\cup$ $F_{new}$\;   
\caption{Algorithm: Find all functions accessible from the processing phase.
}
\label{dynsta:algo}
\end{algorithm}

\section{Multithreaded
Applications}
\label{app:imp}
\sysname uses LLVM's \texttt{lli} interpreter to interpret
the applications. The \texttt{lli} interpreter is a single-threaded
application and does
not support the interpretation of multithreading
applications.
Therefore, to support the analysis of multithreaded applications,
\sysname 
interprets \emph{only} the main thread till the transition point
and treats all child threads as part of the processing phase. 
In other words, \emph{all} code in the child threads will always be
statically analyzed.

The child threads are usually created
using well-known interfaces such as \texttt{pthread\_create} which
accept a
function argument that acts as the thread \emph{entry-point}. 
We modify \texttt{lli}'s handling of function calls to these interfaces 
to record the thread entry-point specified in such calls. 
When partitioning the code, as described in 
Section~\ref{design:partition}, we add all such thread entry-point
functions to the initial list of functions accessible from the
processing phase. Therefore, this approach ensures soundness but potentially
loses precision because all child threads are always completely 
statically analyzed. Adding multithreaded capabilities to the
\texttt{lli} interpreter would mitigate this potential loss of
precision. We leave this task for
future work.


\section{Ablation Study}
\label{app:ablation}
\begin{figure*}
    \centering
    \includegraphics[width=0.90\textwidth]{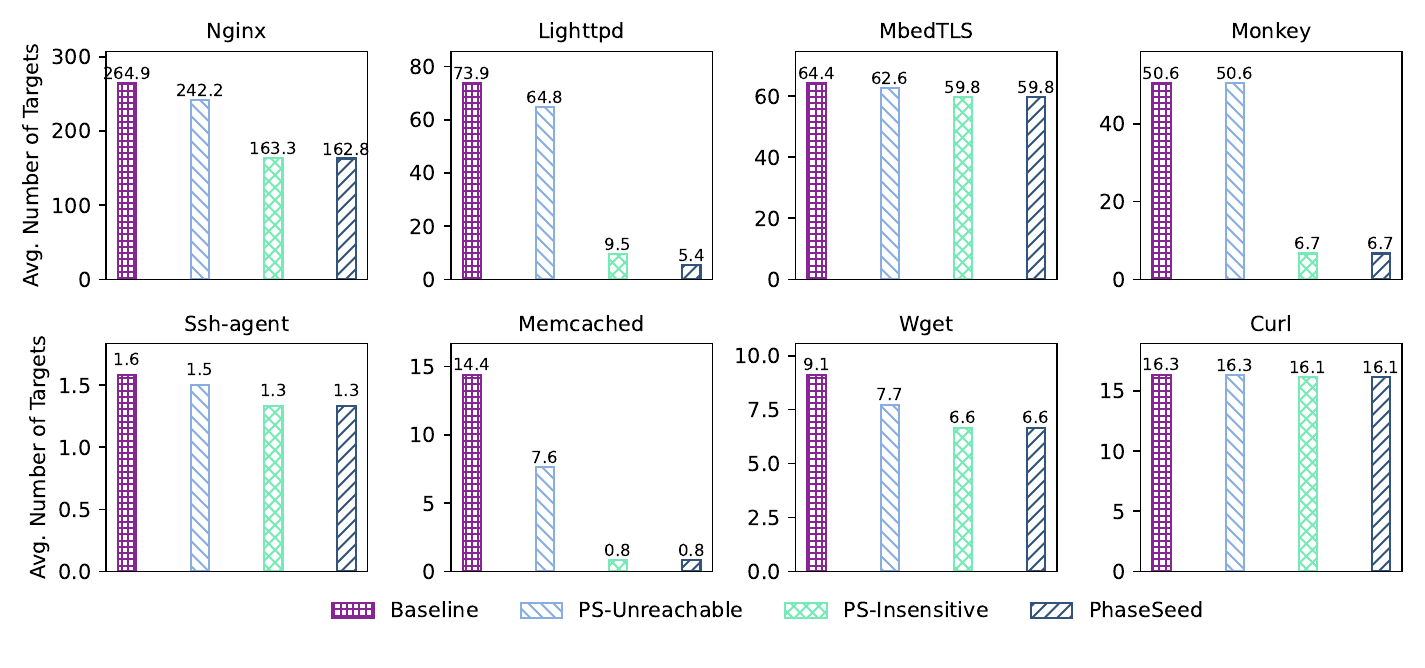}
    \caption{Ablation Study: Average number of CFI targets.}
    \label{fig:avg_cfi_ablation}
\end{figure*}

\begin{figure*}
    \centering
    \includegraphics[width=0.90\textwidth]{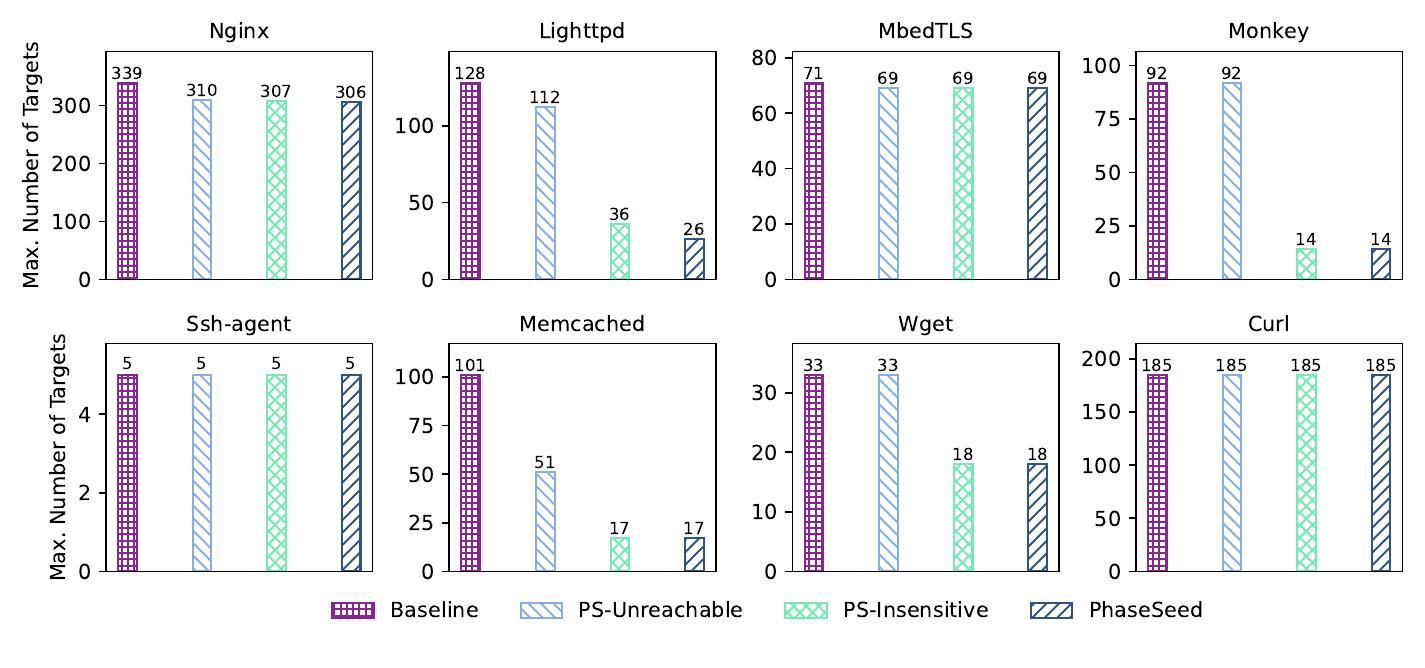}
    \caption{Ablation Study: Maximum number of CFI targets}
    \label{fig:max_cfi_ablation}
\end{figure*}

In this section, we present the results of the ablation study
where we study the precision improvement impact 
of each component of \sysname for the Control Flow Integrity use case. 
We perform the ablation study
with three configurations, in addition to the baseline system,
and report both the average and the maximum sizes of equivalence
classes.

\para{Configurations.} The first configuration, \emph{PS-Unreachable},
isolates the precision improvement
due to the removal of the code that becomes unreachable because of
the runtime configuration disabling certain code paths,
as discussed in \autoref{sec:unreachable}. The second
configuration, \emph{PS-Insensitive}, disables
the fully-sensitive dynamic execution discussed in \autoref{sec:fullysensitive} and only captures
the effect of the compounding precision improvement. 
In this configuration, the points-to heap relationships
are not captured in a context-sensitive manner, at the end
of the dynamic execution. Finally,
the third configuration, \emph{PhaseSeed}, enables
the fully-sensitive dynamic execution and captures the total
precision improvement due to all components of our system.

\para{Effect of Unreachable Code.} As shown in \autoref{fig:avg_cfi_ablation} and 
\autoref{fig:max_cfi_ablation}, in the case for all applications,
except Curl, disabling unreachable code alone does not account
for the total precision improvement provided by \sysname. For 
example, in the case of Nginx, filtering the unreachable code
alone reduces the average EC size from 
264.9 to 242.2. But the compounding precision improvement
and fully-sensitive dynamic execution reduces the average number
of targets to 162.8.

\para{Fully-sensitive Dynamic Execution.}
The fully sensitive dynamic execution shows
significant precision benefits only for Lighttpd where it
reduces the average EC size from 9.5 to 5.4. We analyzed the
results and found that in the case of 
Nginx, we found that while the fully sensitive dynamic
execution provided initial improvements in precision, those
precision gains were almost completely lost by subsequent 
imprecision in the static analysis stage. In the case of Monkey,
ssh-agent, Memcached, and Wget, they compounding precision
improvements already accounted for most of the possible
precision improvements. As discussed in \autoref{eval:cfi},
Curl and MbedTLS show minimal improvement under \sysname
due to their program structure.

This shows that the different components of \sysname
offers different degrees of precision
improvement depending on the application's structure.

\section{Security Critical System Calls}
\label{app:security-critical}
To evaluate the protection that system call filters
can realistically provide, earlier research first 
defined a list of ``security-critical'' system calls
by systematically mining large repositories of 
real-world exploit payloads.
Temporal Specialization~\cite{temporal} extracted 53
Metasploit and 514 Shellstorm payloads and, after
adding equivalent calls, analyzed 1,726 payloads. 
In that corpus, \texttt{bind} and \texttt{select} 
were the most common, appearing 316 and 293 times,
respectively; this metric underlies later 
syscall-filtering evaluations~\cite{syspart}.
We use this data-set for our system call evaluation 
in Section \ref{sec:syscall}.

\section{Analysis Time}
\label{app:analysistime}
\autoref{tbl:analysistime} presents the total analysis time
for both the baseline SVF analysis and the complete \sysname
toolchain. Across all applications, \sysname requires less
analysis time than the baseline fully 
static pointer analysis technique.
This is because the baseline analysis 
spends multiple iterations deriving imprecise pointer relationships, 
whereas \sysname reaches the fixpoint faster. This shows
that \sysname not only improves the analysis precision but
also improves its scalability by reducing the analysis time.

\renewcommand{\arraystretch}{1.05}
\begin{table}[t]
    \centering
    \caption{Analysis time for the baseline and \sysname}
    \label{tbl:analysistime}
    \scalebox{0.90}{
        \begin{tabular}{lrr}
        \toprule
        & \multicolumn{2}{c}{\textbf{Analysis Time (in seconds)}} \\
        \midrule
\bfseries Application &  \bfseries SVF &
\bfseries \sysname \\
        \midrule
Nginx	&	1385.74	&	1382.96	\\
Lighttpd	&	190.41	&	136.84	\\
Mbedtls	&	16.15	&	8.41	\\
Monkey	&	114.14	&	31.64	\\
Ssh-agent	&	16.75	&	10.27	\\
Memcached	&	72.68	&	43.01	\\
Wget	&	32.12	&	12.03	\\
Curl	&	911.81	&	511.32	\\
\bottomrule
\end{tabular}
}
\end{table}